\documentclass[fleqn,twoside]{article}
\usepackage{espcrc2}
\usepackage{amssymb}

\usepackage{graphicx}
\usepackage{epsfig}
\newcommand{\AmS}{{\protect\the\textfont2
  A\kern-.1667em\lower.5ex\hbox{M}\kern-.125emS}}

\newcommand{\mon}{{\mathrm{mon}}}

\newcommand{\be}{\begin{equation}}
\newcommand{\ee}{\end{equation}}
\newcommand{\beqn}{\begin{eqnarray}}
\newcommand{\eeqn}{\end{eqnarray}}
\newcommand{\eq}[1]{(\ref{#1})}

\title{
\thispagestyle{empty}
\vspace{-25mm}
\rightline{\small KANAZAWA-03-17~~~~~}
\rightline{\small ITEP-LAT-2003-12~~~~~}
\vspace{10mm}
Entropy of monopoles from percolating cluster
in quenched SU(2) QCD\thanks{Presented by K. I. at Lattice'03.}}

\author{Katsuya~Ishiguro\address{Institute for Theoretical Physics, Kanazawa University,
Kanazawa 920-1192, Japan},
M. N. Chernodub${}^{\mathrm{a,}}$\address{ITEP, B.Cheremushkinskaya 25, Moscow,
117259, Russia}\thanks{M.N.Ch. is supported by JSPS Fellowship P01023.},
Katsuya~Kobayashi${}^{\mathrm{a}}$
and Tsuneo Suzuki${}^{\mathrm{a}}$\thanks{T.S. is partially
supported by JSPS Grant-in-Aid for Scientific Research on Priority Areas
No.13135210 and (B) No.15340073. This work is also supported by the
Supercomputer Project of the Institute of Physical and Chemical
Research (RIKEN).  A part of our numerical simulations have been done
using NEC SX-5 at Research Center for Nuclear Physics (RCNP) of Osaka
University.}
}
\begin{document}

\begin{abstract}
The length distribution and the monopole action of the infrared monopole clusters
are studied numerically in quenched SU(2) QCD. We determine the effective entropy of the monopole
currents which turns out to be a descending function of the blocking scale,
indicating that the effective degrees of freedom of the extended monopoles are getting smaller
as the blocking scale increases.
\end{abstract}

\maketitle

\section{INTRODUCTION}

The dual superconductor picture~\cite{DualSuperconductor} of the QCD vacuum
is based on the existence of Abelian monopoles which appear naturally in an
Abelian gauge. There are various indications that the mo\-no\-po\-les are responsible
for the color confinement (for a review, see Ref.~\cite{Reviews}). One of
the most important results is the observation of the monopole
condensate in the confinement phase~\cite{shiba:condensation,MonopoleCondensation}.

The monopole trajectories form clusters. Typically, each lattice configuration
contains many finite-sized (ultraviolet) clusters and one large percolating (infrared)
cluster~\cite{ivanenko,ref:kitahara,ref:hart}.
The IR cluster -- which occupies the whole lattice -- represents the monopole condensate~\cite{ivanenko}.
The tension of the confining string gets a dominant contribution from the
%monopoles belonging to the
IR cluster~\cite{ref:kitahara}. In the deconfinement phase the IR cluster
disappears~\cite{ivanenko,ref:kitahara}, as expected.

The UV and IR monopole clusters were studied previously in
Refs.~\cite{ref:kitahara,ref:hart,ref:boyko,ref:zakharov:clusters,zakharov:recent}.
Below we investigate the action and length distribution of the infrared monopole cluster
for various lattice sizes and scales $b$ at which the magnetic charge is defined.
Using the action and length distribution we investigate the entropy of the IR clusters.

\section{MODEL}

In our simulations we use the Wilson action, $S(U) = - \frac{\beta}{2}\, {\mathrm{Tr}} U_P$.
We work in the Maximal Abelian (MA) gauge~\cite{kronfeld}. The
Abelian gauge field, $\theta$, is determined as a phase the diagonal component of the $SU(2)$ link
variable, $\theta_\mu(s) = \arg\, U^{11}_{\mu}(s)$. The Abelian field
strength, $\theta_{\mu\nu}(s)\in(-4\pi,4\pi)$, is decomposed into two parts,
$\theta_{\mu\nu}(s)= \bar{\theta}_{\mu\nu}(s) +2\pi m_{\mu\nu}(s)$,
where $\bar{\theta}_{\mu\nu}(s)\in [-\pi,\pi)$ is interpreted as
the electromagnetic flux through the plaquette
and $m_{\mu\nu}(s)$ can be regarded as a number of the Dirac
strings piercing the plaquette.

The elementary monopole currents are
defined as follows~\cite{degrand}, $ k_{\mu}(s) =1/2 \epsilon_{\mu\nu\rho\sigma}
\partial_{\nu}m_{\rho\sigma}(s+\hat{\mu})$,
where $\partial$ is the forward lattice derivative.
To study the monopole charges at various scales we use the
$n^3$ extended monopole construction~\cite{ivanenko},
\beqn
k_{\mu}^{(n)}(s) = \!\!\sum_{i,j,l=0}^{n-1}\!\!k_{\mu}(n s+(n-1)\hat{\mu}+i\hat{\nu}
     +j\hat{\rho}+l\hat{\sigma}).\nonumber
\eeqn
The extended monopole is defined
on a sublattice with the spacing $b=na$, where $a$ is
the spacing of the original lattice. Both elementary and extended monopole charges
are conserved and quantized.

We used the standard Monte--Carlo procedure to generate 1000-3000 configurations of the gauge field
for each value of $\beta=2.1 \sim 2.6$. To fix the MA gauge we used either the usual iterative
algorithm (for $24^4$ and $48^4$ lattices) or the
Simulated Annealing method~\cite{ref:bali} with five Gribov copies (for $32^4$ lattice).

\section{MONOPOLE ACTION}
\label{sec:action}

To get the monopole action we integrate out all degrees of freedom but
the monopole ones. Following Ref.~\cite{shiba:condensation} we generate the SU(2)
configurations, fix the MA gauge, then get the configurations of
the IR monopole clusters as it was described in the previous section.
Then we use the inverse Monte--Carlo method to get the monopole action.

The monopole action can be represented~\cite{shiba:condensation,chernodub} as a sum of
the $n$--point ($n \ge 2$) operators $S_i$:
$$
S_{\mon}[k] = \sum_i f_i S_i [k]\,,
\vspace{-2mm}
%\label{eq:monopole:action}
$$
where $f_i$ are the coupling constants. We adopt only the two--point interactions
in the monopole action ($i.e.$ interactions of the form $S_i \sim k_{\mu}(s) k_{\mu'}(s')$).
A detailed description of the $S_i[k]$ interactions can be found in Ref.~\cite{IshiguroSuzuki}.
\vskip -6mm
\begin{figure}[h]
\centerline{\includegraphics[angle=0,scale=0.24,clip=true]{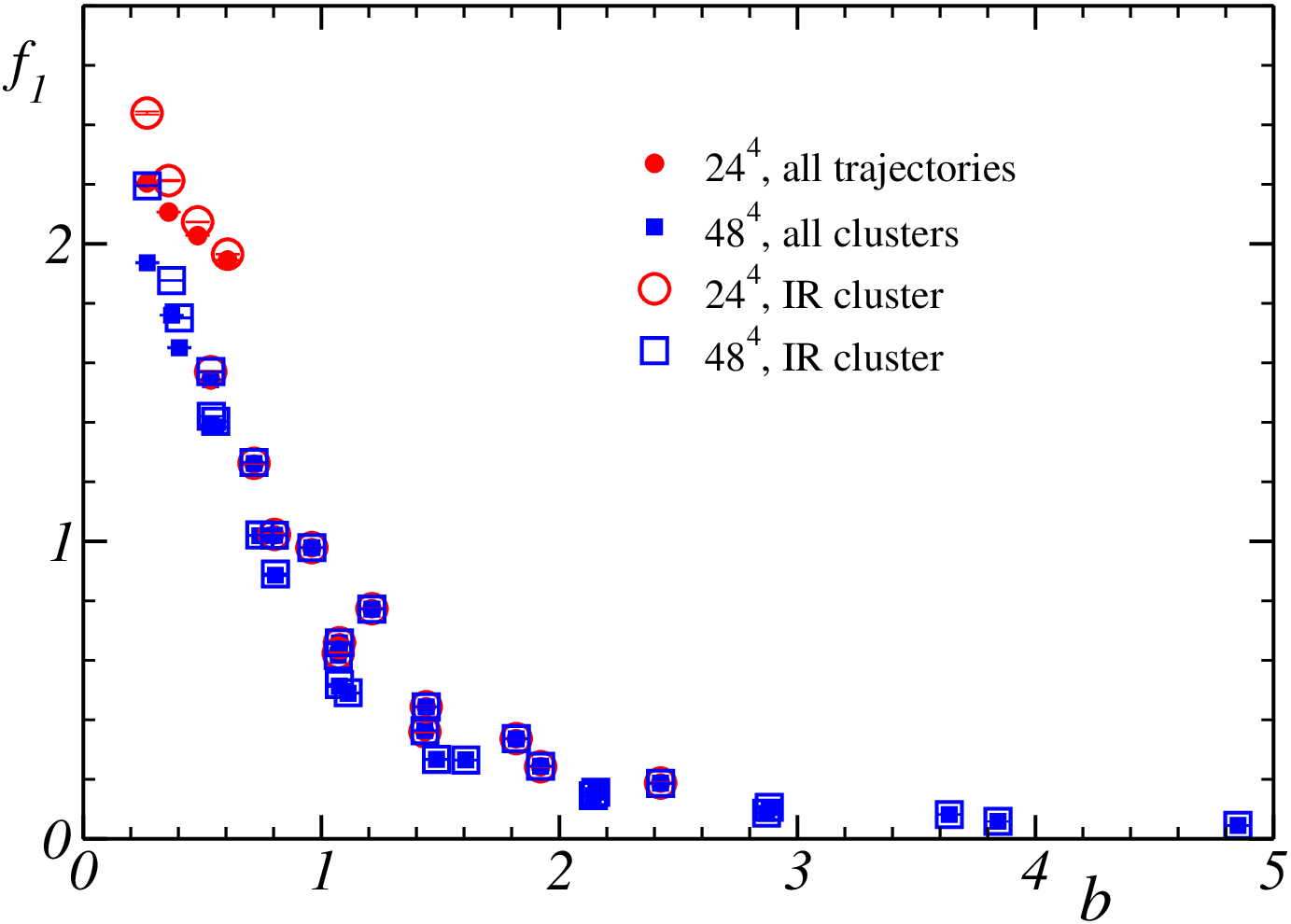}}
\vskip -8mm
\caption{The self--interaction coupling $f_1$ $vs.$ $b$.}
\label{fig:f1}
\end{figure}
\vskip -7mm

One can find that the monopole action is proportional with a good accuracy to the length
of the monopole loop, $S_{\mon} = f L + const$. The dominant term in the monopole action
corresponds to the most local self-interaction of the monopole currents,
$S_1[k] = \sum_{s,\mu} k^2_\mu(s)$. We compare the parameter $f_1$ for the monopole action of
the IR cluster and for the action associated with the whole monopole ensemble
in Figure~\ref{fig:f1}.

One can notice that the coupling constant $f_1$ shows scaling for large $b=na$ in agreement
with Ref.~\cite{shiba:condensation}. This coupling is independent of the lattice volume
and for large blocking scales $b$ the type of the ensemble (the IR cluster or the
whole ensemble) is not essential for determination of $f_1$. However, at small $b$ values,
$b \sqrt{\sigma} \lesssim 0.5$, the type of the lattice ensemble becomes important.

\section{LENGTH DISTRIBUTION}
\label{sec:length}

The distribution of the ultraviolet clusters was studied
both numerically~\cite{ref:hart,zakharov:recent} and analytically~\cite{ref:zakharov:clusters,ref:zakharov}.
The distribution can be described by a power law $D_{UV} \propto L^{-\tau}$,
where the power $\tau$ is very close to 3, Ref.~\cite{ref:hart}. This behaviour
indicates that the monopoles in UV clusters are randomly walking objects~\cite{ref:zakharov:clusters}.
In our simulations we are concentrated on the IR cluster because
the IR cluster is important for the confinement of quarks.

The length distribution function, $D(L)$, is proportional to the
weight with which the particular trajectory of the length $L$ contributes to the partition function.
From the previous section it is clear that monopole action contributes to $D(L)$
in a form of an exponential factor, $\propto e^{- f L}$. The entropy of the monopole trajectory also
contributes to the monopole length distribution. The entropy contribution is proportional
to $\mu^L$ (with $\mu>0$) for
sufficiently large monopole lengths, $L$. Thus the distribution of the monopole trajectories in
infinite volume must be described by the function
\beqn
D^{IR}_{\inf}(L) \propto \mu^L \cdot e^{ - f \, L} = e^{\gamma L}\,, \quad \gamma = \ln \mu - f\,.
\label{eq:IR:distr:inf}
\eeqn
In this equation we neglect a power-law prefactor which is essential for the distribution of the
ultraviolet clusters. In the finite volume there appears a finite--volume
suppression factor and the total distribution can be described as a Gaussian~\cite{IshiguroSuzuki}:
\beqn
D^{IR}(L) = const. \, \exp\{ - \alpha(b,V) L^2 + \gamma(b) L\}\,.
\label{eq:IR:distr:two}
\eeqn
The peak of this distribution, $L_{max} = \gamma(b) \slash 2 \, \alpha(b,V)$,
is expected to be proportional to the volume of the system, $V$, to insure
the finiteness of the IR monopole density, $\rho_{IR} = L_{max}/ V$,
in the thermodynamic limit, $V \to \infty$. Thus we conclude that
$\alpha(b,V) = A(b) \slash V$, where $A(b)$ is a certain function. Therefore
in the thermodynamic limit the parameter $\alpha$ vanishes and the
distribution \eq{eq:IR:distr:two} is reduced to Eq.\eq{eq:IR:distr:inf}, as expected.

\begin{figure}[thb]
\centerline{\includegraphics[angle=0,scale=0.24,clip=true]{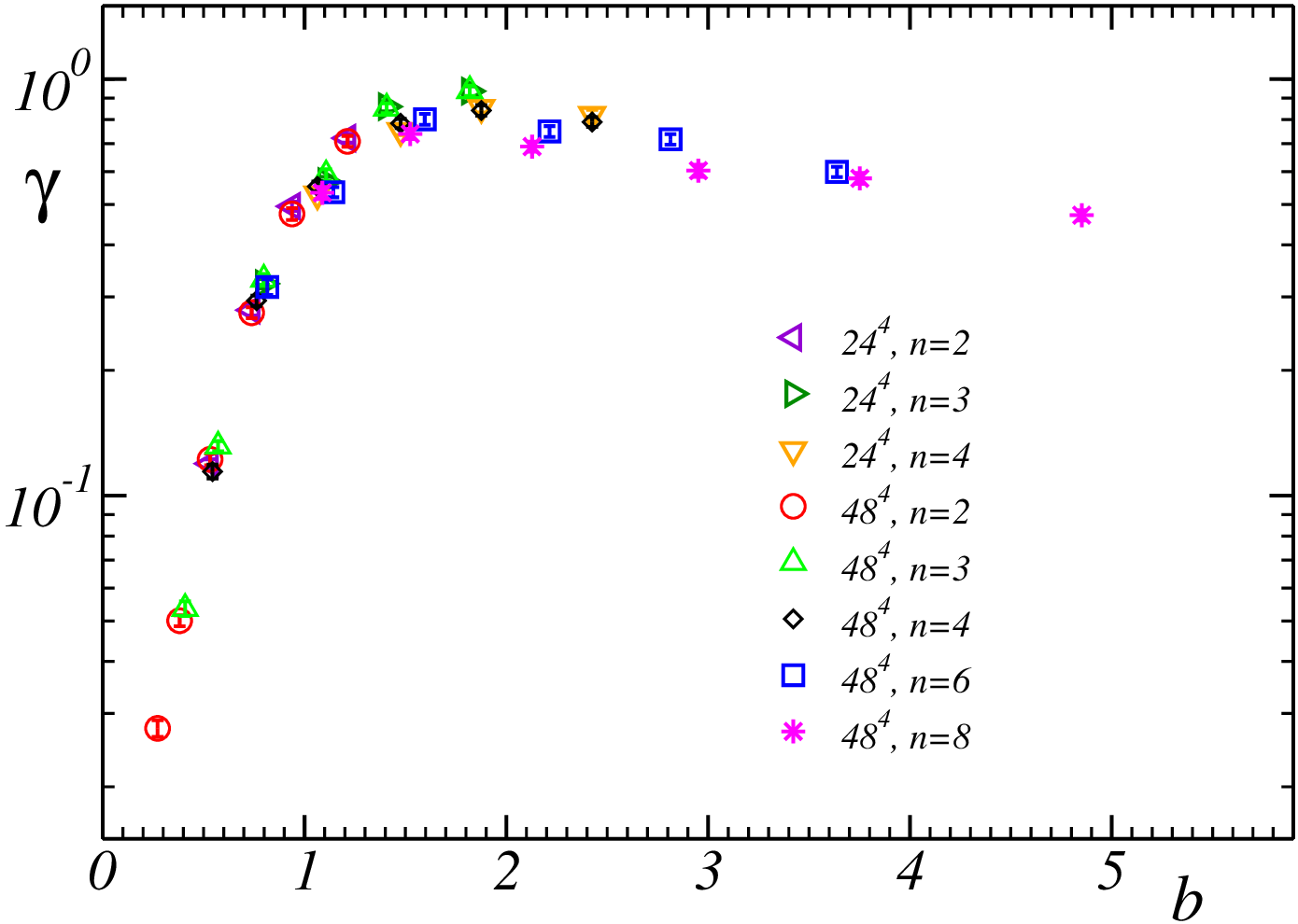}}
\vskip -6mm
\caption{The fitting parameter $\gamma$ as functions $b$ for
various lattice volumes and blocking steps, $n$.}
\label{fig:alpha:gamma}
\vskip -6mm
\end{figure}

To get the monopole entropy we need to know the parameter $\gamma$, Eq.~\eq{eq:IR:distr:inf}.
We fit the monopole loops distributions by the function~\eq{eq:IR:distr:two} and use
the bootstrap method to estimate the statistical errors. We obtain that the parameter $\gamma$
--- shown in Figure~\ref{fig:alpha:gamma} --- scales with  $b=n\cdot a$ and is independent
on the volume of the system.

\section{MONOPOLE ENTROPY}

The knowledge of the distribution and
the monopole action allows us to determine the entropy of the monopole currents.
According to Eq.~\eq{eq:IR:distr:inf}, $\mu = e^{\gamma + f}$.
If the monopoles make a simple random walk on a $4D$ hypercubic lattice
then we get $\mu=7$ since there are seven choices
at each site for the monopole current to go further.

\begin{figure}[t]
\centerline{\includegraphics[angle=0,scale=0.24,clip=true]{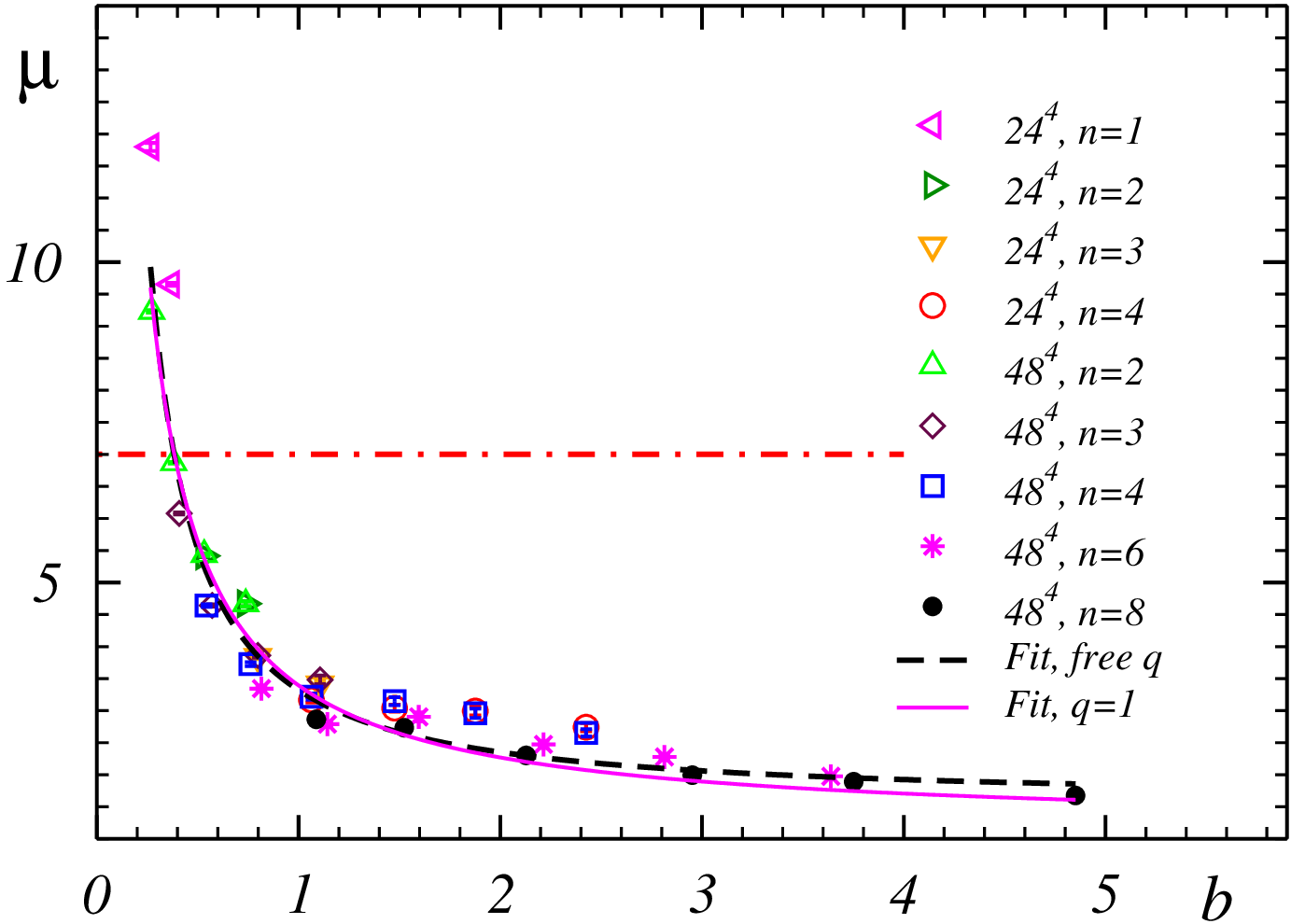}}
\vskip -6mm
\caption{Entropy factor $\mu$ $vs.$ $b$. The lines represent the fits by Eq.~\eq{eq:fit:entropy}.}
\label{fig:entropy}
\vskip -6mm
\end{figure}

The entropy factor $\mu$ is shown in
Figure~\ref{fig:entropy}. It scales with $b$ and is independent of the volume
of the lattice. Note that in a small $b$ region $\mu>7$ because the monopole action can not be
reliably described by the quadratic terms only~\cite{chernodub}.

At large $b$ the entropy factor is smaller than seven. We have fitted the entropy by
%
%the
a
 function:
\beqn
%\vspace{-2mm}
\mu^{\mathrm{fit}} = \mu_{\infty} + C\, \mu^{-q}\,,
\label{eq:fit:entropy}
%\vspace{-2mm}
\eeqn
where $\mu_{\infty}$, $C$ and $q$ are the fitting parameters. The fit gives
$\mu_{\infty} = 1.6(4)$, $C = 1.7(5)$ and $q = 1.2(2)$. Fixing $q=1$
in Eq.~\eq{eq:fit:entropy} we get $\mu_{\infty} = 1.15(25)$ and $C = 2.2(1)$.
The corresponding best fit curves are shown
in Figure~\ref{fig:entropy}.

The fact that the asymptotic value of the entropy is very close to the unity in the
large $b$ limit may have a simple explanation. A monopole with a
large blocking size $b$ behaves as a classical object and its motion is no more a simple
random walk. The predominant motion of the large--$b$ monopole is close to a straight line.

\end{document}